\documentclass[linenumbers]{aastex631}

\usepackage{amsmath}

\newcommand{\anglB}[1]{\left(#1\right)}
\newcommand{\squB}[1]{\left[#1\right]}
\newcommand{\curlB}[1]{\left\{ #1\right\}}
\newcommand{\absB}[1]{\left| #1 \right|}

\newcommand{\vprint}[1]{\textbf{\textit{#1}}}
\newcommand{\rbent}{r_\mathrm{f}}
\newcommand{\rout}{r_\mathrm{max}}
\newcommand{\thmax}{\theta_\mathrm{max}}

\begin{document}
\nolinenumbers
\title{Thermal Spectra of Warped and Broken Accretion Disks}

\author[0000-0001-9239-0373]{J. Speicher}\thanks{E-mail: jspeicher3@gatech.edu}
\affiliation{Center for Relativistic Astrophysics, School of Physics, Georgia Institute of Technology, 837 State Street, Atlanta, GA 30332-0430, USA}

\author[0000-0002-8082-4573]{O. Blaes}
\affiliation{Department of Physics, University of California, Santa Barbara, CA 93106, USA}

\begin{abstract}
    Black holes may accrete gas with angular momentum vectors misaligned 
    with the black hole spin axis. The resulting 
    accretion disks are subject to Lense-Thirring precession, and hence torque. Analytical calculations and simulations show that Lense-Thirring precession will warp, and, for large misalignments, fracture the disk. In GRMHD simulations, the warping or breaking occurs at $\lesssim10r_s$, where $r_s$ is the Schwarzschild radius. Considering that accretion disk spectra in the soft state of stellar-mass black holes are generally well modeled as multicolor blackbodies, the question arises as to how consistent warped and broken disks are with observations. Here, we analytically calculate thermal spectra of warped and broken disks with a warp or break radius at $10r_s$ for various disk inclinations. Due to self-irradiation and the projected area of the inclined disk regions, the spectra of inclined disks significantly deviate from multicolor blackbodies and do not follow the multicolor blackbody relation $\nu L_\nu\propto\nu^{\gamma}=\nu^{4/3}$ at low frequencies $\nu$. The power-law indices at low frequencies of the inclined disks vary with viewing angle; when viewed face-on, they vary between $\gamma\approx0.91-1.26$ for the warped disks and $\gamma\approx1.37-1.54$ for the broken disks depending on the inclination angle. The differences decrease when moving the location of the disk warp and break to larger radii; for inclined disks to emit as multicolor blackbodies, they must warp or break at radii $\geq50r_s$. Our results imply that accretion disks around black holes in the soft state warp or break at larger radii than suggested in GRHMD simulations. 
\end{abstract}

\section{Introduction} \label{sec:intro}


Accretion disks are ubiquitous in nature, forming around objects ranging from young stars to neutron stars up to supermassive black holes \citep[e.g.,][]{Williams2011ARA&A..49...67W,Alma2015ApJ...808L...3A,EHT2019ApJ...875L...1E,Eijnden2020MNRAS.493.1318V}. For black holes, accretion disks allow the accreted material to move inwards while transporting angular momentum outwards through turbulence (such as that arising from the magnetorotational instability), magnetocentrifugal winds, and possibly spiral waves
\citep[e.g.,][]{Blandford1982MNRAS.199..883B,Balbus1998RvMP...70....1B,Spruit1989ASIC..290..325S,MunozDarias2019ApJ...879L...4M}. 

The angular momentum vector of the accreted material can misalign with the black hole spin axis. The misalignment can arise for instance through natal kicks during supernovae \citep[e.g.,][]{Kalogera2000ApJ...541..319K,Gerosa2013PhRvD..87j4028G}, or due to turbulent accretion during galaxy mergers \citep[][]{King2006MNRAS.373L..90K}. Observationally, misalignment of accretion disks has been indicated by precessing jets and from gravitational wave measurements   
\citep[e.g.,][]{Hjellming1995Natur.375..464H,Orosz1997ApJ...477..876O,OShaughnessy2017PhRvL.119a1101O,Cui2023Natur.621..711C}.

The misalignment of the black hole spin axis and the angular momentum axis of the accretion flow will impact the accretion flow structure \citep[for a review on tilted accretion disks, see, e.g.,][]{Fragile2024arXiv240410052F}. An inclined disk around a spinning black hole experiences Lense-Thirring precession, which induces torque on the disk. In the calculations by \cite{Bardeen1975ApJ...195L..65B}, Lense-Thirring precession aligns the accretion disk with the black hole spin at small radii, and the disk smoothly transitions into an inclined disk at larger radii. However, the torque due to Lense-Thirring precession increases with inclination angle, and the disk can fracture when it surpasses viscous torques \citep{Nixon2012ApJ...757L..24N}. Smoothed particle hydrodynamics simulations found disk tearing to occur for inclinations $\gtrsim45^\circ$ \citep[e.g.,][]{Nelson2000MNRAS.315..570N,Nixon2012ApJ...757L..24N,Nealon2015MNRAS.448.1526N}. Disks breaking at inclinations $\gtrsim45^\circ$ have also been found in GRMHD simulations, with the difference that the disk warp or break occurs closer to the black hole \citep[e.g.,][]{Liska2019MNRAS.487..550L,Liska2021MNRAS.507..983L,Liska2022ApJS..263...26L}. 


Given the significant impact of Lense-Thirring precession on the accretion disk structure in simulations, the question arises as to how the inclined accretion disk structure translates into observational properties. 
The soft state of a stellar-mass black hole is dominated by thermal emission, which is thought to originate from a geometrically thin but optically thick accretion disk \citep[e.g.,][]{Dotani1997ApJ...485L..87D,Frontera2001ApJ...546.1027F}. If the disk is misaligned with the black hole spin axis, this emission should come from a warped or broken disk. Nevertheless, black hole systems with high orbital inclinations and spins can be adequately modeled as a multitemperature, color-corrected 
blackbody, with the blackbody temperature varying with radius \citep[e.g.,][]{Makishima1986ApJ...308..635M}, emitted by a flat disk \citep[e.g.,][]{Gierlinski2004MNRAS.347..885G,Mall2024MNRAS.52712053M}. 
If the disks around these black hole systems are warped or broken, their emission should therefore resemble the emission of an inclined flat disk. 
In this work, we test the inclined disk geometry of GRMHD simulations and compute the emission of warped and broken disks around stellar-mass black holes and compare them with flat-disk multicolor blackbody spectra. For this comparison, the paper is structured as follows. The calculations of the disk emission are described in section \ref{sec:methods} and presented in section \ref{sec:results}. We discuss our results in section \ref{sec:discussion} and conclude with section \ref{sec:conclusion}.


\section{Methods}\label{sec:methods}
\subsection{Disk Geometry}\label{subsec:geometry}
We consider infinitely thin disks extending from the radius of the innermost stable circular orbit $r_{\mathrm{isco}} = r_s$, corresponding to a high black hole spin parameter of $a_*\approx0.95$, to a maximum radius $\rout=10^5r_s$. The radius $r_s$ is the Schwarzschild radius,
\begin{equation}
    r_s = \frac{2GM}{c^2},
\end{equation}
where $G$ is the gravitational constant, $c$ is the speed of light, and $M$ is the black hole mass set to $10M_\odot$. 

In the first considered setup, the accretion disk is warped, and its inclination increases smoothly with radius (section \ref{subsubsec:warpedgeo}). In the second setup, the accretion disk is broken, consisting of a flat inner region and an inclined flat outer disk region (section \ref{subsubsec:brokengeo}). We set both the warp radius, defined as the ratio of the Lense-Thirring precession frequency and the effective kinematic 
viscosity in the vertical direction \citep{Scheuer1996MNRAS.282..291S}, and the radius separating the flat and inclined region of the broken disk to $\rbent=10r_s$. This choice places the warp and break similarly close to the black holes as in GRMHD simulations \citep[e.g.,][]{Liska2021MNRAS.507..983L,Liska2022ApJS..263...26L}. 

We describe our inclined disks using spherical coordinates $\curlB{r,\theta,\phi}$. 
Angles in the frame of the disk have a $\sim$ superscript. Angles in the disk and in the unrotated frame are the same in uninclined 
disk regions. 


\subsubsection{Warped Disk Geometry}\label{subsubsec:warpedgeo}
The unit angular momentum vector $\vprint{l} = \curlB{l_x,l_y,l_z}$ of a steady-state, warped accretion disk with constant viscosity is \citep{Scheuer1996MNRAS.282..291S},
\begin{equation} 
\begin{split}
l_x & = \sin\thmax\cos\anglB{2\sqrt{\frac{\rbent}{r}}} \exp\anglB{-2\sqrt{\frac{\rbent}{r}}} \\
 l_y &= \sin\thmax\sin\anglB{2\sqrt{\frac{\rbent}{r}}} \exp\anglB{-2\sqrt{\frac{\rbent}{r}}}\\
l_z &=  \sqrt{1-\anglB{l_x^2 + l_y^2}}.
\end{split}\label{eq:lxlylz}
\end{equation}
leading to a radially dependent disk inclination $i$ of,
\begin{equation}
    i = \cos^{-1}l_z,\label{eq:diskInclination}
\end{equation}
and disk coordinates in the disk frame
\begin{equation}
\begin{aligned}
    x &=  r\cos\tilde{\phi}\\
    y &= r\sin\tilde{\phi}\cos i\\
    z &= r\sin\tilde{\phi} \sin i.\label{eq:cartesianDisk}
\end{aligned}\end{equation}
Eq.\ref{eq:cartesianDisk} neglects the twisting of the disk implicit in eq.\ref{eq:lxlylz}; the twist is mainly present at small radii (where the inclination
is in any case small) and will thus not majorly impact the implications of the warping at larger radii.

Setting eq.\ref{eq:cartesianDisk} equal to the standard definition of spherical coordinates, $\curlB{x,y,z} = \curlB{r\sin\theta\cos\phi, r\sin\theta\sin\phi,r \cos\theta}$, allows solving for $\tilde{\phi}$ in terms of unrotated angles,
\begin{equation}
    \begin{aligned}
    \tilde{\phi} &= \arctan\squB{\frac{\tan\phi}{\cos i}},
\end{aligned}\end{equation}
so that the components of the disk position vector $\vprint{r}=\curlB{x,y,z}$ can be reexpressed as 
\begin{equation}
\begin{aligned}
    x &=  r\cos\anglB{\arctan\squB{\frac{\tan\phi}{\cos i}}} \\
    y &= r\sin\anglB{\arctan\squB{\frac{\tan\phi}{\cos i}}}\cos i \\
    z &= r\sin\anglB{\arctan\squB{\frac{\tan\phi}{\cos i}}} \sin i.\label{eq:carthesianActual}
\end{aligned}\end{equation}


\begin{figure}
    \centering
    \includegraphics[width = 0.8\textwidth]{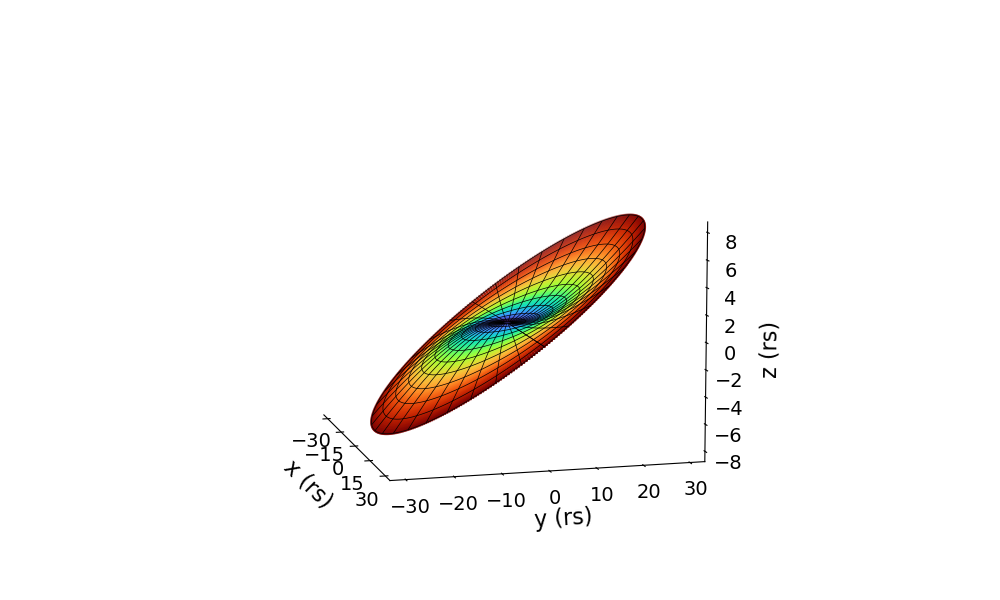}
    \caption{The inner $r\leq30r_s$ of a warped disk with a maximum inclination of $\thmax=0.3\pi$. The color changes for visualization.}
    \label{fig:warpedDisk}
\end{figure}
Fig. \ref{fig:warpedDisk} shows an example of a warped disk with $\thmax=0.3\pi$. We plot $r\leq30r_s$ and vary the color with radius for visualization. Our parameterization (eq.\ref{eq:carthesianActual}) yields a disk symmetric along the y-axis and antisymmetric along the x-axis.  

An area element $dA$ of a warped disk is
\begin{equation}
    dA = \sqrt{\absB{\vprint{v}_r\times \vprint{v}_\phi}}\,dr\,d\phi,\label{eq:dawr}
\end{equation}
where $\vprint{v}_r$ and $\vprint{v}_\phi$ are nonnormalized tangent vectors pointing in the radial and the angular direction respectively, 
\begin{equation}
   \begin{aligned}
    \vprint{v}_r &= \frac{\partial\vprint{r}}{\partial r} \\
    \vprint{v}_\phi &= \frac{\partial\vprint{r}}{\partial \phi} 
    \label{eq:tangentVec}.
\end{aligned}
\end{equation}

The unit normal vector $\vprint{n}$\footnote{The vector $\vprint{n}$ (eq.\ref{eq:n}) approaches $\vprint{l}$ (eq.\ref{eq:lxlylz}) rotated by $\pi/2$ as $r\rightarrow\infty$ and $\phi=\pi/2$. The different pointing direction is due to a different disk orientation and does not impact the results.} is the normalized cross product of $\vprint{v}_r$ and $\vprint{v}_\phi$,
\begin{equation}
    \vprint{n} = \frac{\vprint{v}_r\times \vprint{v}_\phi}{\absB{\vprint{v}_r}\absB{\vprint{v}_\phi}}.\label{eq:n}
\end{equation}

\subsubsection{Broken Disk Geometry}\label{subsubsec:brokengeo}
The broken disk inclination is discontinuous at $\rbent$,
\begin{equation}
     i = \left\{\begin{matrix}
0 & r\leq \rbent \\
\thmax & r > \rbent \\
\end{matrix}\right. .\label{eq:inclBroken}
\end{equation}

\begin{figure}
    \centering
    \includegraphics[width = 0.8\textwidth]{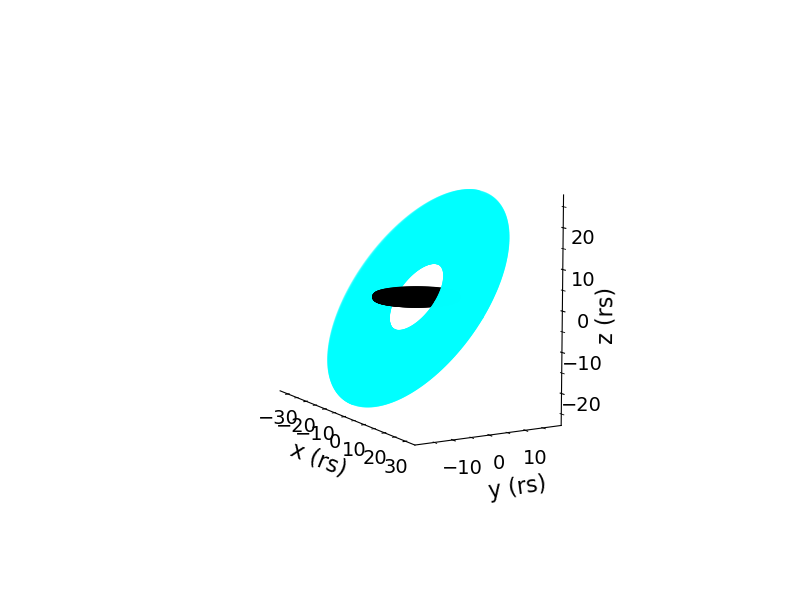}
    \caption{A broken disk setup, where the disk is flat within $r\leq\rbent$ (black region) and inclined by $\thmax=0.3\pi$ otherwise (cyan region). For better visualization, we only plot the region $r\leq30r_s$.}
    \label{fig:brokenDisk}
\end{figure}
The region within $r\leq30r_s$ of a broken disk is shown in Fig. \ref{fig:brokenDisk}. Beyond $\rbent$ (black region), the disk is rotated by $\thmax=0.3\pi$ around the x-axis (cyan region). The rotation by an arbitrary angle $\alpha$ around the x-axis can be described with a rotation matrix $\mathcal{R}_\alpha$,
\begin{equation}
    \mathcal{R}_\alpha=  \begin{bmatrix}
1 & 0 & 0 \\
0 & \cos\anglB{\alpha} & -\sin\anglB{\alpha} \\
0 & \sin\anglB{\alpha} & \cos\anglB{\alpha} \\
\end{bmatrix}.\label{eq:rot}
\end{equation}
For simplicity, we will call the rotation matrix (eq.\ref{eq:rot}) by $\thmax$ $\mathcal{R}$. 
The disk coordinates in the unrotated frame therefore are,
\begin{equation}
    \vprint{r} = \left\{\begin{matrix}
r\curlB{\cos\phi,\sin\phi,0} & r\leq \rbent \\
\mathcal{R}\,r\curlB{\cos\tilde{\phi},\sin\tilde{\phi},0} & r > \rbent \\
\end{matrix}\right.,\label{eq:r}
\end{equation}

and the normal vector $\vprint{n}$ is
\begin{equation}
    \vprint{n} = \left\{\begin{matrix}
\curlB{0,0,1} & r\leq \rbent \\
\mathcal{R}\,\curlB{0,0,1} & r > \rbent \\
\end{matrix}\right..\label{eq:nbr}
\end{equation}
The area element $dA$ in the accretion disk frame is 
\begin{equation}
    dA = r\,d\Tilde{\phi}\,dr.\label{eq:dabr}
\end{equation}.

\subsection{Disk temperature}\label{subsec:diskTemperature}

For a standard accretion disk extending down to the $r_{\mathrm{isco}}$, the viscous heating rate is \citep{Shakura1973A&A....24..337S}
\begin{equation}
   Q_v =  \frac{3 G M \dot{M}}{8\pi r^3}\anglB{1-\sqrt{\frac{r_\mathrm{isco}}{r}}}.\label{eq:qv}
\end{equation}
We set the mass accretion rate $\dot{M}$ to $0.05\dot{M}_{Edd}$, where $\dot{M}_{Edd}$ is the Eddington mass accretion rate and related to the Eddington luminosity $L_{Edd}$ as $\dot{M}_{Edd} = L_{Edd}/\anglB{0.1 c^2}$. For pure hydrogen accretion, $\dot{M}\approx7\times10^{17}$ gs\textsuperscript{-1}.

In standard accretion disk theory, integrating eq.\ref{eq:qv} over the entire accretion disk gives the disk luminosity
\begin{equation}
    L_{d} = \int_0^{2\pi}\int_{r_\mathrm{isco}}^\infty Q_v \,dA = \frac{GM\dot{M}}{2r_\mathrm{isco}}, \label{eq:ldisktheory}
\end{equation}
and is $\approx1.6\times10^{38}$ erg s\textsuperscript{-1} for our setup. Evaluating eq.\ref{eq:ldisktheory} between $r_\mathrm{isco}$ and $\rout$ for our inclined disks yields deviation from the theoretical value of $\lesssim0.2\%$, proving that $\rout$ is sufficiently large (the spatial integration is described in section \ref{subsec:numericalEvaluation})


\begin{figure}
    \centering
    \includegraphics[width=0.45\linewidth]{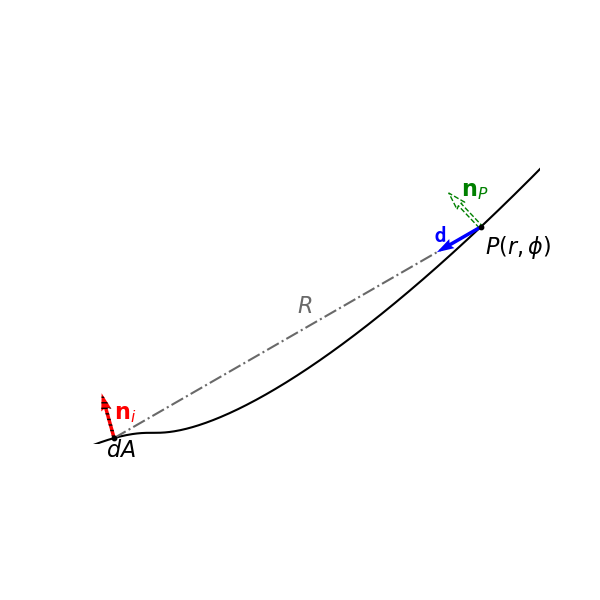}
    \includegraphics[width=0.45\linewidth]{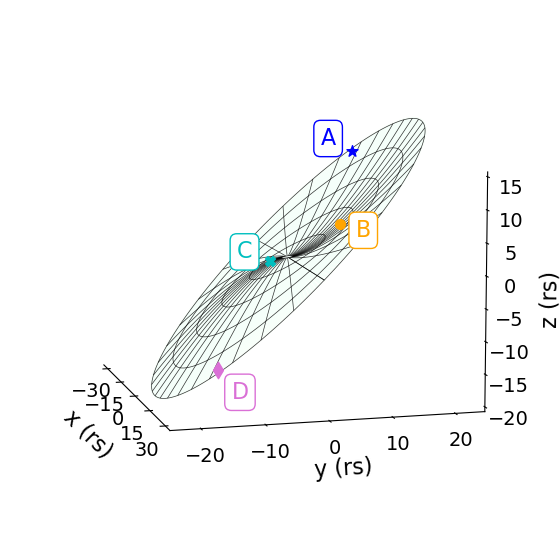}
    \vspace{-15mm}
    \caption{Left panel: Irradiation of $P(r,\phi)$ by disk element $dA$ \citep[see also][]{Fukue1992PASJ...44..663F}. The point $P(r,\phi)$ and $dA$ are a distance $R$ (grey dot-dashed line) apart. The unit vector $\vprint{d}$ (blue vector) points from $P(r,\phi)$ to $dA$. The normal vectors at $P(r,\phi)$ and $dA$ are $\vprint{n}_P$ (green dashed vector) and $\vprint{n}_i$ (red striped vector) respectively. Right panel: Illustration of the visibility between points on the upper warped disk surface. Points on the concave part on the disk (points A, blue star, and B, orange dot) can irradiate each other. Points on the convex part (points C, cyan cross, and D, pink diamond) can not irradiate each other. A point $P_1\anglB{r,0<\phi_1<\pi}$ (e.g., points A and B) can irradiate another point $P_2\anglB{r,\pi<\phi_2<2\pi}$ (e.g., points C and D) and vice versa if the angle between the line connecting the points and the normal vector at $P_2$ is less than $\pi/2$ (satisfied for B and C, but not for B and D).}
    \label{fig:irradiationSetup}
\end{figure}

A warped or broken disk will irradiate itself \citep[left panel of Fig. \ref{fig:irradiationSetup} for a warped disk; see also][]{Fukue1992PASJ...44..663F}. An infinitesimal disk element $dA$ (eq.\ref{eq:dawr} or \ref{eq:dabr} depending on the disk setup) with a local temperature $T$ irradiates a point $P(r,\phi)$ with an intensity $I = \sigma T^4/\pi$, assuming the disk radiates as a blackbody everywhere. The unit vector \vprint{d} (blue vector) points from $P(r,\phi)$ to $dA$, which is a distance $R$ (grey dot-dashed line) away. The normal vectors at $P(r,\phi)$ and $dA$ are called $\vprint{n}_{\mathrm{P}}$ (green dashed vector) and $\vprint{n}_{\mathrm{i}}$ (red striped vector) respectively. The heating due to irradiation $dQ_{irr}$ of $P(r,\phi)$ due to $dA$ is \citep[see also][]{Fukue1992PASJ...44..663F},
\begin{equation}
    dQ_{irr} = \frac{I}{R^2}\absB{\anglB{-\vprint{d}\cdot\vprint{n}_{\mathrm{i}}}\anglB{\vprint{d}\cdot\vprint{n}_{P}}\anglB{\vprint{n}_{P}\cdot\vprint{n}_{\mathrm{i}}}}\,dA.\label{eq:dQirr}
\end{equation}
Eq. \ref{eq:dQirr} assumes no albedo. While the albedo will be nonzero in the inner disk regions, we are focusing on the impact of the inclined outer regions, where a zero albedo is a reasonable assumption \citep[e.g.,][]{Taverna2020MNRAS.493.4960T}. 
We also neglect light bending, which would increase the irradiation of the inner disk region and reduce the amount of irradiation at large radii due to the emission from small radii. However, light bending would mostly impact the emission originating close to the black hole, and would therefore not significantly contribute to any deviations from the flat disk solution due to the disk inclination at large radii.

We integrate $dQ_{irr}$ (eq.\ref{eq:dQirr}) over the entire accretion disk (the spatial integration is described in section \ref{subsec:numericalEvaluation}). Any point $P(r,\phi)$ will only be irradiated by area elements in its line of sight (right panel of Fig. \ref{fig:irradiationSetup}). The top surface of the warped disk between $0<\phi<\pi$ is concave and convex between $\pi<\phi<2\pi$. Two points between $0<\phi<\pi$ (e.g., points A, blue star, and B, orange dot) can thus always irradiate each other and between $\pi<\phi<2\pi$ (e.g., points C, cyan cross, and D, pink diamond) never irradiate each other . A point $P_1\anglB{r_1,0<\phi_1<\pi}$ can irradiate another point $P_2\anglB{r_2,\pi<\phi_2<2\pi}$ and vice versa if the angle between $\overrightarrow{P_1P_2}$ and $\vprint{n}_{P_1}$ is less than $\pi/2$ (Point B and C, but not B and D). The integral of eq.\ref{eq:dQirr} gives the irradiation onto the upper warped disk surface at $P(r,\phi)$. Due to symmetry, evaluating eq.\ref{eq:dQirr} at $P(r,\phi+\pi)$ yields the bottom irradiation at $P(r,\phi)$. 

Two points on the broken disk can irradiate each other if they are not on the same plane. The absolute value of the dot products in eq.\ref{eq:dQirr} accounts for both disk surfaces, so integrating eq.\ref{eq:dQirr} gives the total heating due to irradiation $Q_{irr}$ for the broken disk at $P(r,\tilde{\phi})$. 

The total heating due to irradiation at point $P(r,\phi)$ is thus
\begin{equation}
     Q_{irr}\bigg\rvert_{P( r,\phi)} = \left\{\begin{matrix}
    \int dQ_{irr}\bigg\rvert_{P( r,\phi)}+\int dQ_{irr}\bigg\rvert_{P(r,\phi+\pi)}
 &  \mathrm{warped\,disk}\\
\int dQ_{irr}\bigg\rvert_{P( r,\tilde{\phi})} &  \mathrm{broken\,disk}\\
\end{matrix}\right. .\label{eq:qint}
\end{equation}
Due to the disk symmetries (Fig. \ref{fig:warpedDisk}, \ref{fig:brokenDisk}), $Q_{irr}$ (eq.\ref{eq:qint}) is the same for all quadrants. To sample the whole disk, we calculate $Q_{irr}$ for 24 angles between 0 and $\pi/2$ and for 24 angles between $\pi$ and $3\pi/2$ at different radii.

Self-irradiation heats the disk, increasing the emission intensity. We therefore iterate the $Q_{irr}$ calculations until convergence. The disk temperature is initially set by viscous heating (eq.\ref{eq:qv}), and then by viscous heating and self-irradiation, $T = \squB{\anglB{Q_v+Q_{irr}}/2\sigma}^{1/4}$, in the subsequent iterations. 
The inner disk region is the greatest source of radiation, but its temperatures change negligibly through self-irradiation due to the comparatively low temperatures at larger radii. Heating through self-irradiation therefore converges quickly, and we stop after the third iteration, which changed $Q_{irr}$ by less than $5\%$ for all $\thmax$.

\begin{figure*}
    \centering
    \includegraphics[width=0.8\linewidth]{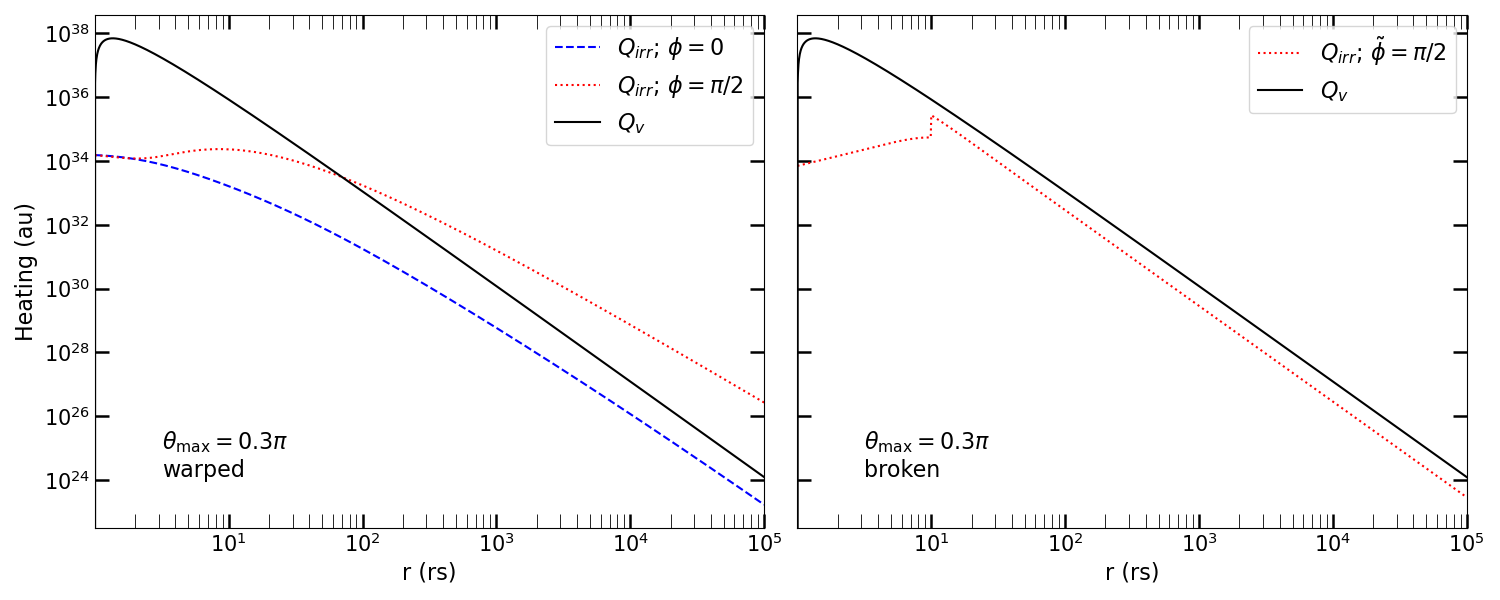}
    \caption{Heating of warped (left panel) and broken (right panel) disks with $\thmax=0.3\pi$. The accretion disks are heated through viscous heating (black lines, eq. \ref{eq:qv}), and self-irradiation (eq. \ref{eq:qint}). We iterated eq. \ref{eq:qint} three times to account for the self-irradiation increasing the irradiating intensity when calculating $Q_{irr}$. Self-irradiation is radially and angle dependent, decreasing with radius and being at a maximum at $\phi=\pi/2$ (red dotted lines) and a minimum at $\phi=0$ (blue dashed line in the left panel, zero for a broken disk and thus not plotted).}
    \label{fig:heating}
\end{figure*}
Fig. \ref{fig:heating} shows $Q_v$ (black solid lines, eq.\ref{eq:qv}) and $Q_{irr}$ for $\phi = \pi/2$ and $\phi=0$ (eq.\ref{eq:qint}) for a warped (left panel) and a broken disk (right panel) with $\thmax=0.3\pi$. For the broken disk (right panel), $Q_{irr}$ is measured at $\Tilde{\phi}$. The heating due to self-irradiation reaches a minimum at $\phi=0$ (blue dashed line in the left panel, zero for the broken disk and thus not plotted) and $\phi=\pi$ and a maximum at $\phi=\pi/2$ (red dotted lines) and $\phi=3\pi/2$ for either disk setup.  At small radii, $Q_{irr}$ at $\phi=\pi/2$ are similar for both inclined disks and significantly smaller than $Q_v$. At $2r_s$ for instance, $Q_v$ is $\sim 3800$ times greater than $Q_{irr,\phi=\pi/2}$ of the warped disk and $\sim3200$ times greater than for the broken disk. The hot inner disk region will emit radiation at high energies; we therefore expect $Q_v$ to mostly drive the high-energy emission.

However, $Q_{irr}$ of a warped disk decreases slower with radius than $Q_v$ and therefore becomes increasingly important at larger radii. After an increase of $Q_{irr}$ at $\rbent$ for $\phi=\pi/2$ (left panel in Fig. \ref{fig:heating}, red dotted line), $Q_{irr,\phi=\pi/2}$ surpasses $Q_v$ at $r\approx70r_s$. Since the warped disk is approximately flat at $\phi=0$, $Q_{irr,\phi=0}$ (left panel, blue dashed line) does not increase at $\rbent$. With $Q_{irr}$ dominating over $Q_v$ in the cooler disk regions, self-irradiation is expected to noticeably impact the low-energy emission of warped disks.


For broken disks, $Q_{irr}$ at $\Tilde{\phi}=\pi/2$ (right panel in Fig. \ref{fig:heating}, red dotted line) increases abruptly at $\rbent$ due to the disk discontinuity. Beyond $\rbent$, $Q_{irr,\Tilde{\phi}=\pi/2}$ decreases like $Q_v$; fitting a power-law to the heating rates for $r\geq \rbent$ yields a radial dependence of about $r^{-3}$ for both. Since $Q_{irr}$ is always $\gtrsim3$ times smaller than $Q_v$ for a broken disk, self-irradiation will be less important than for warped disks. Nevertheless, the geometrical aspect of the inclined disk region will still impact the spectrum.


\subsection{Disk spectrum}\label{subsec:diskspec}
The accretion disk spectrum as a function of frequency $\nu$ is calculated as 
\begin{equation}
  \nu L_\nu = \frac{2h\nu^4}{c^2}
  \int \frac{\absB{cos \beta}\,dA}{f_{c}^4\squB{\exp\anglB{h\nu/kf_cT}-1}},  \label{eq:nuFnu}
\end{equation}

where $k$ is the Boltzmann constant and $h$ the Planck constant. We interpolate $Q_{irr}$ (eq.\ref{eq:qint}) with radii and angles to obtain disk temperatures for every angular and radial increment of the spatial integral (section \ref{subsec:numericalEvaluation}).

The color correction factor $f_c$ is \citep[][]{Done2012MNRAS.420.1848D},
\begin{equation}
     f_{c} = \left\{\begin{matrix}1 & T_{tot}< 3\times 10^4\,\mathrm{K} \\
\left ( \frac{T_{tot}}{3\times 10^4\,\mathrm{K}} \right )^{0.82} &3\times 10^4\,\mathrm{K}\leq  T_{tot}\lesssim 10^5\,\mathrm{K} \\
\left ( \frac{72\,\mathrm{keV}}{kT_{tot}} \right )^{1/9}  & T_{tot}\gtrsim 10^5\,\mathrm{K}  \\\end{matrix}\right. ,\label{eq:fc}
\end{equation}
and accounts for scattering effects. 

The angle $\cos\beta$ is the dot product of the normal vector $\vprint{n}$ (eq. \ref{eq:n} or \ref{eq:nbr}, depending on the disk setup) and a unit vector pointing towards the observer. For an observer with viewing angle $\mu$, 
\begin{equation}
    \cos\beta = \vprint{n}\cdot\mathcal{R}_\mu\curlB{0,0,1},\label{eq:cosbeta}
\end{equation}
with $R_\mu$ set by eq.\ref{eq:rot}.

We use eq.\ref{eq:nuFnu} to also test for energy conservation by varying $\cos\beta$ not only in the $\theta$, but also in the $\phi$ direction and integrating over all solid angles. The calculated luminosity integrated over all solid angles deviates from the theoretically expected value (eq.\ref{eq:ldisktheory}) by $<10$\%.


Only disk portions visible to the observer contribute to eq.\ref{eq:nuFnu}. For $\mu \leq\pi/2$, all warped disk regions are visible. For $\mu>\pi/2$, the innermost region of the warped disk can be shadowed, and an observer may see parts of the top and bottom surface. Extracting the visible regions of the top surface for viewing vectors $R_\mu \curlB{0,0,1}$ and $-R_\mu \curlB{0,0,1}$ gives the total visible surface for $\mu > \pi/2$. A point $P(r,\pi < \phi<2\pi)$ on the top surface of the warped disk is invisible if the angle between its normal vector and the viewing vector is greater than $\pi/2$. To determine whether a point $P_1(r_1,0 < \phi_1<\pi)$ is shadowed by some other point $P_2(r_2,\pi < \phi_2<2\pi)$, we first conduct a binomial search for a 720-entry long angle interval between  $\pi < \phi<2\pi$; for every $\phi$, we minimize the dot product between normal and viewing vector to find the radius $r_{v}$ beyond which the disk is visible to an observer. For a point $P_1(r_1,0 < \phi_1<\pi)$, we then find the corresponding point $P_{v}\anglB{r_{v},\phi_{v}}$ in the observer's line of sight. If the angle between the viewing vector and the horizontal plane is greater than the angle between the horizontal plane and the line $\overline{P_{v}P_1}$, $P_1$ is not visible. To check the visibility of broken disk regions, we collapse the disk onto a two-dimensional plane as seen by the observer. We compare the collapsed positions of the flat region with the collapsed points of the most and least inclined positions, and the collapsed position of the inclined region with the collapsed points of the outermost flat regions to determine whether a point is shadowed. The resulting spectrum is evaluated for a logarithmically spaced 960-entry long frequency grid ranging from $10^{-4}$keV$/h\approx2.4\times10^{13}$ Hz to $20$ keV$/h\approx4.8\times10^{18}$ Hz. The spatial integration is described in section \ref{subsec:numericalEvaluation}.




We calculate spectra for a 20-entry long $\mu$ interval ranging between 0 and $0.95\pi$ of warped and broken disks with $\thmax = 0.1\pi$, 0.2$\pi$, 0.3$\pi$, and $0.4\pi$. To explore the impact of $\rbent$, we also consider warped and broken disks with $\rbent=50r_s$, $500r_s$ and $1000r_s$ with a final inclination of $0.3\pi$ at $\rout$, which requires increasing $\thmax$ to $0.321\pi$ if $\rbent=50r_s$, to $0.38\pi$ if $\rbent=500r_s$, and to $0.44\pi$ if $\rbent=1000r_s$ for the warped disk setup. For another warped disk setup with a final inclination of $0.2\pi$ at $\rout$ and $\rbent=500 r_s$, $\thmax=0.236\pi$. We will state $\rbent$ explicitly for these additional disk configurations. Otherwise, $\rbent=10r_s$. Since $Q_v$, $Q_{irr}$ and thus also the disk spectra depend on $\dot{M}$, we also calculated warped disk spectra with $\rbent=10r_s$, $\thmax=0.3\pi$, and $\dot{M}=0.02\dot{M}_{Edd}$. Unless specified otherwise, $\dot{M}=0.05\dot{M}_{Edd}$, though.

\subsection{Spatial integral evaluation}\label{subsec:numericalEvaluation}

For warped disks, all integrations are evaluated in the unrotated frame. For the broken disk setup, the integrations are performed in the tilted frame, i.e., using $\anglB{r,\tilde{\phi}}$. However, we switch to the unrotated frame when necessary (e.g., for calculating the vectors and $R$ in eq.\ref{eq:dQirr}). 

The viscous heating rate (eq.\ref{eq:qv}, Fig.\ref{fig:heating}) decreases quickly with radius. Hence, we perform an adaptive integration of eq. \ref{eq:ldisktheory} to split the radial integral into subintegrals with increasing ranges and step sizes. To avoid $Q_v=0$ at $r_\mathrm{isco}$, the first subintegral starts at $\anglB{1+10^{-5}}r_s$ with a radial step size of $\approx3.7\times10^{-5}r_s$, ending at $\approx\anglB{1+4\times10^{-4}}r_s$. The radial step sizes increase up to $625 r_s$ for the last radial integral from $\approx93750r_s$ to $10^5r_s$. 

For the broken disks, the angular integrations in eq. \ref{eq:ldisktheory}, \ref{eq:qint}, and \ref{eq:nuFnu} use a 720-entry long angle interval ranging from 0 to $2\pi$. For the warped disks, eq. \ref{eq:ldisktheory} and \ref{eq:qint} use a 720-entry long angle interval ranging from 0 to $2\pi$ as well, but we exploit the disk symmetry for eq. \ref{eq:nuFnu} with two 192-entry long angle intervals between $0$ and $\pi/2$ and between $3/2\pi$ and $2\pi$ and double the angular integral results to cover all angles; choosing the smaller interval ranges changes the result by less than 1 \%. All integrals are calculated using the Simpson method.




\section{Results}\label{sec:results}
\subsection{Disk Spectra}\label{subsec:discspectra}



\begin{figure*}
    \centering
    \includegraphics[width=0.8\linewidth]{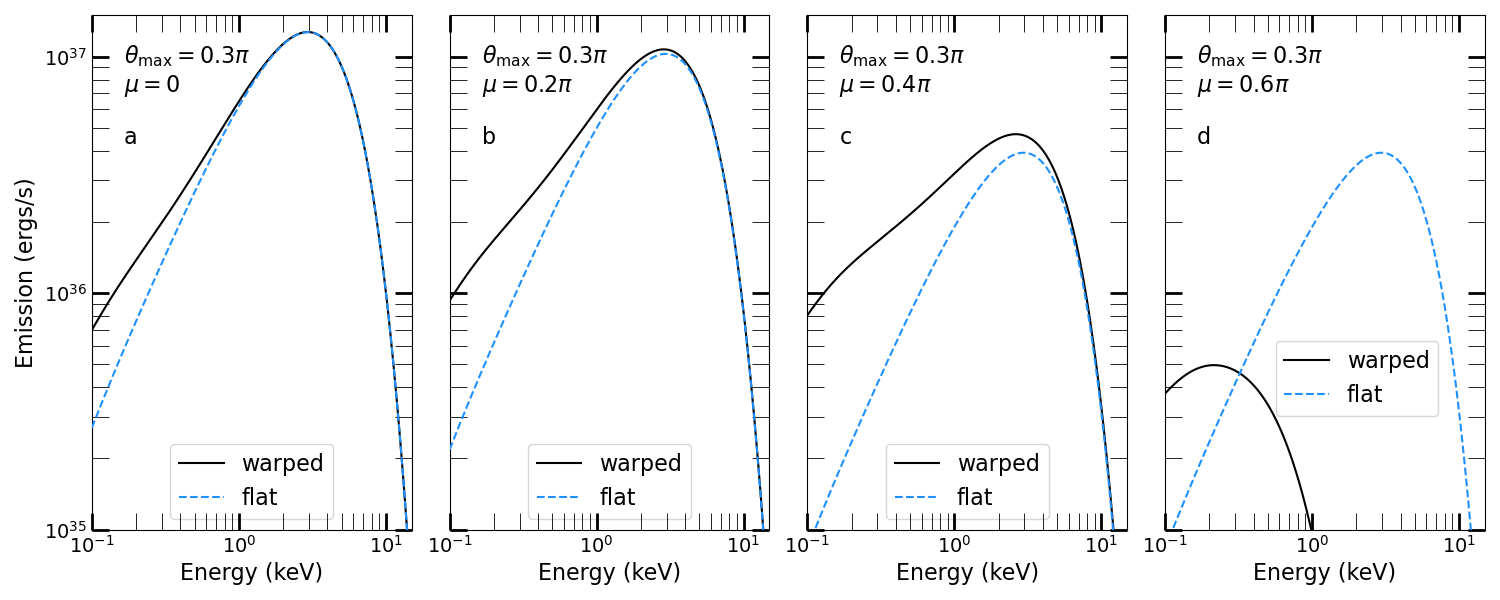}
    \caption{Spectra of the warped disk with $\thmax=0.3\pi$ (black solid lines) and flat disk (blue dashed lines) when viewed at an angle of $\mu=0$ (a), $\mu = 0.2\pi$ (b), and $\mu=0.4\pi$ (c), and $\mu=0.6\pi$ (d). The warped-disk emission at low energies is enhanced due to self-irradiation and a projected area increasing the closer $\mu$ is to $\pi/2$.}
    \label{fig:spectraWarped}
\end{figure*}

Fig. \ref{fig:spectraWarped} shows spectra of a warped disk with $\thmax=0.3\pi$ (black solid lines) and a flat disk (blue dashed lines), viewed face on (a), and at angles $0.2\pi$ (b), $0.4\pi$ (c), and $0.6\pi$ (d). The plotted high-energy emission of the flat and the warped disk agree with each other in panels a-c, because it predominantly originates from the flat inner region where the effects of self-irradiation are unimportant (Fig. \ref{fig:heating}, left panel). If $\mu = 0.6\pi$ (d), the innermost region is not visible and only the outer cooler regions contribute to the observed emission, shifting the spectrum to low energies. The low-energy emission of the warped disk dominates over the flat disk emission for all $\mu$ due to self-irradiation increasing the temperatures at large radii (Fig. \ref{fig:heating}), and because the projected area of the outer warped region increases the closer $\mu$ is to $\pi/2$. 

\begin{figure*}
    \centering
    \includegraphics[width=0.8\linewidth]{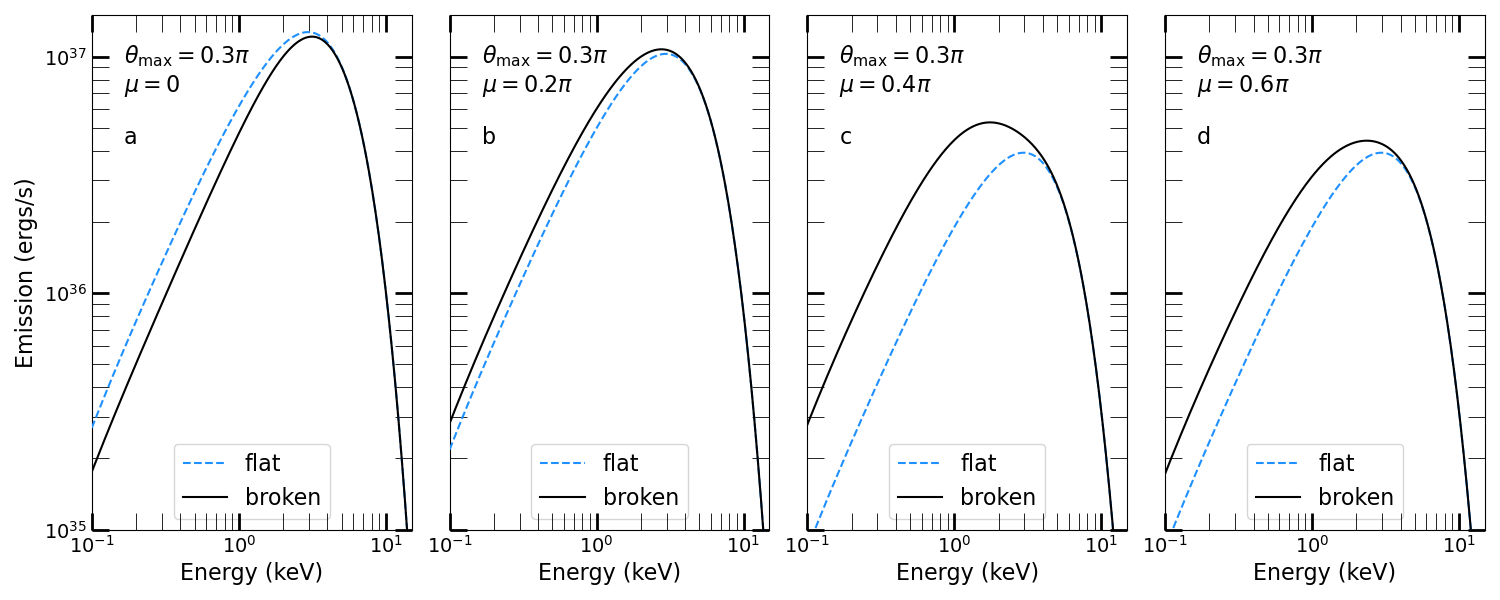}
    \caption{Same as Fig.\ref{fig:spectraWarped}, but for a broken disk with $\thmax=0.3\pi$. The large inclination of the majority of the broken disk region leads to palpable differences between the broken (black solid lines) and flat spectra (blue dotted lines).}
    \label{fig:spectraBroken}
\end{figure*}
Fig. \ref{fig:spectraBroken} shows spectra of a broken disk for the same viewing angles and $\thmax$ as in Fig. \ref{fig:spectraWarped}. Like for the warped disk, the high-energy emission of the broken and flat disk agree with each other. The low energy emission comes largely from the outer inclined region. In contrast to warped disks, the low-energy emission is not significantly enhanced through self-irradiation because $Q_v$ always dominates over $Q_{irr}$ (Fig. \ref{fig:heating}). The low-energy emission $\leq1$ keV of the flat disk is $\approx1.46$ times greater than of the broken disk when $\mu=0$ (a) due to the outer disk being inclined by $\thmax$. In panels b-d, the broken disk emission dominates at low energies because the effective viewing angle of the inclined outer disk is smaller than for the flat disk.



\begin{figure}
    \centering
    \includegraphics[width=0.9\linewidth]{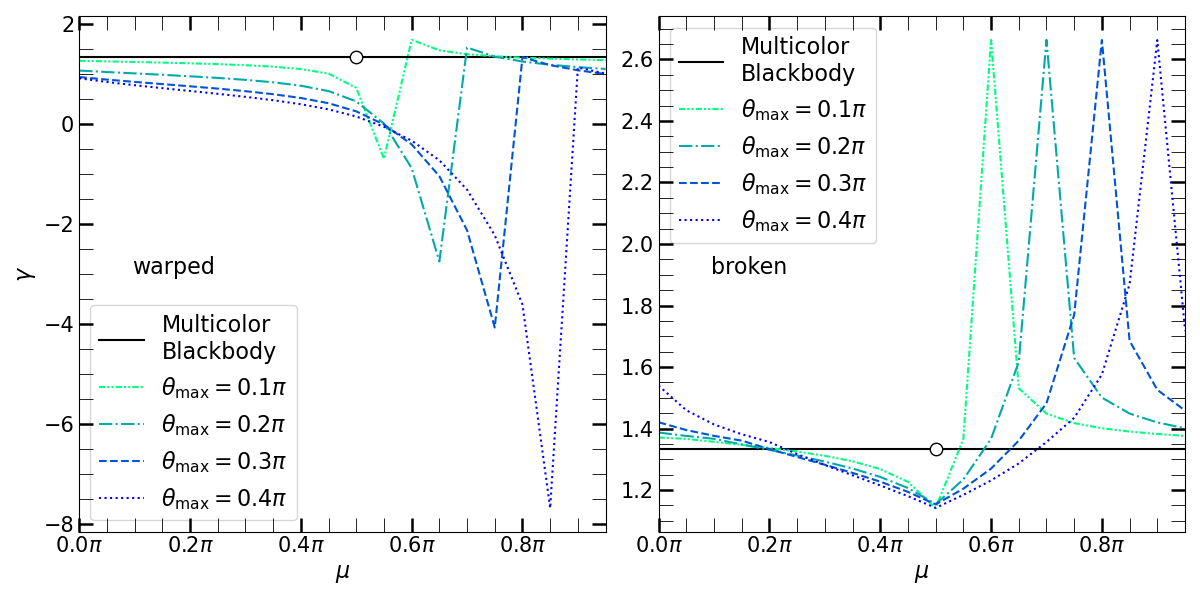}
    \caption{Left panel: The power-law indices of $\nu L_\nu$ between 0.2 keV and 0.5 keV for warped disks for different viewing angles. The indices decrease with decreasing visibility of the innermost disk region and increase again with its regained visibility. Right panel: Same as the left panel, but for broken disks. The indices decrease with increasing $\mu\leq\pi/2$. The sudden increase in $\gamma$ is due to the inclined region being viewed edge-on. The power law indices of the inclined disks significantly differ from the multicolor blackbody solution (black solid lines).}
    \label{fig:expboth}
\end{figure}

Self-irradiation and the projected area of the outer disk region introduces significant differences in the power-law indices at low energies compared to the multicolor blackbody solution (Fig. \ref{fig:expboth}). A multicolor blackbody scales at low energies as $\nu L_\nu\propto\nu^{\gamma}$ with a power-law index of $\gamma=4/3$ \citep[black solid line, e.g.,][]{Makishima1986ApJ...308..635M}. The unfilled circle marks the absence of a multicolor blackbody power-law index at $\mu=\pi/2$, because no emission is observed from an edge-on infintely flat disk. Fitting a power-law to the multicolor blackbodies plotted in Fig. \ref{fig:spectraWarped} and \ref{fig:spectraBroken} in the 0.2-0.5 keV range yields $\gamma\approx1.36$ for all viewing angles, deviating from $\gamma=4/3$ by $\leq2\%$. The power-law indices for that energy range of the warped disk vary between $\gamma\approx0.91-1.26$ at $\mu=0$, with $\gamma$ decreasing with $\thmax$ (left panel). As the inner disk region becomes less visible and the outer disk region more visible, $\gamma$ decreases with increasing $\mu$. The indices become negative when the inner region is completely shadowed and the visible emission peaks at energies $\lesssim0.2$ keV (see rightmost panel in Fig. \ref{fig:spectraWarped}). As $\mu$ increases further, the innermost region is visible again, and $\gamma$ sharply increases before slightly decreasing again.

The power-law indices of broken disks fitted between 0.2 and 0.5 keV decrease for $\mu\leq\pi/2$ in a similar fashion 
(Fig. \ref{fig:expboth}, right panel). Starting with $\gamma\approx1.37-1.54$ at $\mu=0$, the decrease is approximately linear; combining the linear fits for this $\mu$ range gives $\gamma(\theta,\mu \leq \pi/2) = \anglB{a\,\thmax+b}\mu + \anglB{c\,\thmax+d}$, where $a = -0.10\pm0.02$, $b=-0.09\pm0.01$, $c = 0.10\pm0.03$, and $d=1.36\pm0.02$. The inclined region is viewed edge-on at $\mu=\pi/2+\thmax$, raising $\gamma$ to a maximum value of $\approx2.66$ regardless of $\thmax$. The multicolor blackbody solution $\gamma=4/3$ holds in the energy range $kT_{out}\ll E\lesssim0.3k T_{in}$, where $T_{in}$ and $T_{out}$ are the inner and outer disk temperatures respectively \citep{Makishima1986ApJ...308..635M}. Since the visible flat portion only extends out to $\rbent$, the temperatures do not vary enough for that energy range to be applicable. Hence, $\gamma$ significantly differs from the multicolor blackbody solution even when only the flat disk portion is visible.

\begin{figure}
    \centering
    \includegraphics[width=0.8\linewidth]{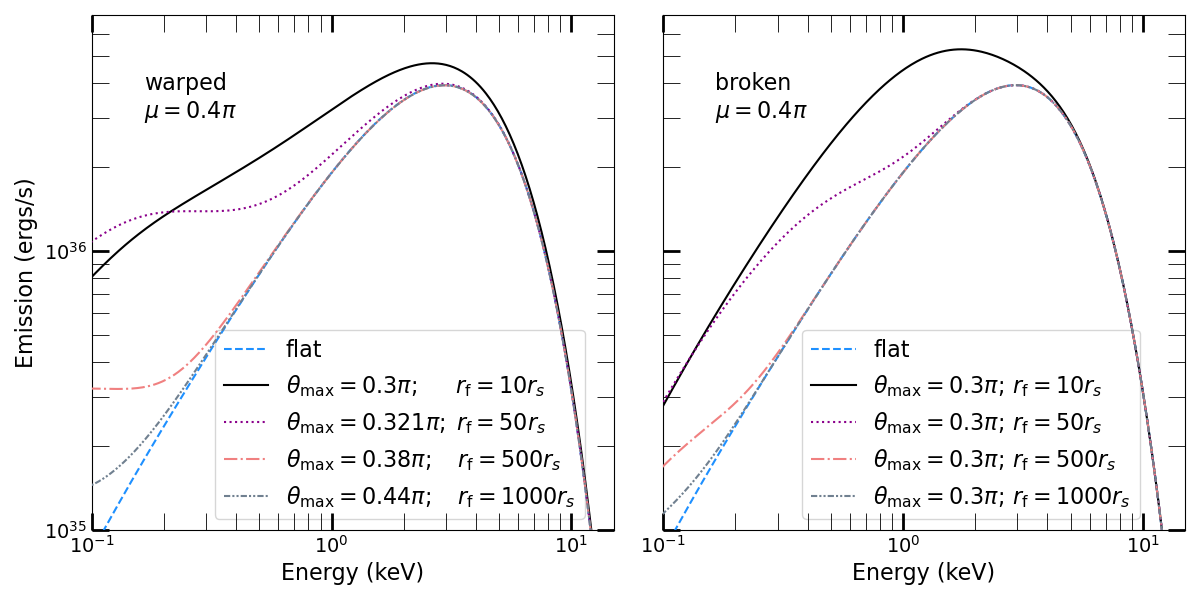}
    \caption{Spectra of warped (left panel) and broken (right panel) disks for $\rbent=10r_s$ (black solid lines, same as in panels c of Fig. \ref{fig:spectraWarped} and \ref{fig:spectraBroken} respectively), $\rbent=50r_s$ (purple dotted lines), $\rbent=500r_s$ (orange dot-dashed lines), and $\rbent=1000r_s$ (grey dash-dotdotted lines) for $\mu=0.4$. The inclination at $\rout$ is $\approx0.3\pi$ for all disks. Increasing $\rbent$ reduces the differences between inclined and flat disk (blue dashed lines) spectra.}
    \label{fig:spectraChangeRW}
\end{figure}

The deviations between inclined and flat disks decrease with increasing $\rbent$. Fig.\ref{fig:spectraChangeRW} shows spectra (eq.\ref{eq:nuFnu}) of warped disks (left panel) and broken disks (right panel) with a maximum inclination of $\approx0.3\pi$ at $\rout$ for $\mu=0.4\pi$ and $\rbent=10r_s$ (black solid lines, same as in panel c of Fig. \ref{fig:spectraWarped} and \ref{fig:spectraBroken} respectively), $\rbent=50r_s$ (purple dotted lines), $\rbent=500r_s$ (orange dot-dashed lines), and $\rbent=1000r_s$ (grey dash-dotdotted lines). Increasing $\rbent$ reduces the projected area of the outer disk regions and confines $Q_{irr}>Q_v$ for warped disks to increasingly larger radii. Hence, the differences at low energies between the inclined and the flat disk spectra (blue dashed lines) decrease. 

The reduction of the low-energy emission is reflected in $\gamma$ fitted between 0.2 keV and 0.5 keV. The power-law indices of the warped disk with $\rbent=10r_s$ (blue dashed line in the left panel of Fig. \ref{fig:expboth}) are significantly lower than the multicolor blackbody solution $\gamma=4/3$ for the majoritiy of viewing angles. While moving $\rbent$ to $50r_s$ and adjusting $\thmax$ to $0.321\pi$ reduces the low-energy emission $\gtrsim0.2$ keV, $\gamma$ does not significantly converge towards the flat disk solution for the calculated viewing angles; $\gamma$ still deviates by $\gtrsim15$\% for the warped disk. Increasing $\rbent$ further to $500r_s$ and $1000r_s$ and increasing $\thmax$ to $0.38\pi$ and $0.44\pi$ respectively moves $\gamma$ towards $4/3$. Except for $\mu=1/2\pi-3/4\pi$, when the innermost region of the warped disk is not visible, $\gamma$ deviates from $4/3$ by $\lesssim31$ \% if $\rbent=500r_s$ and by $\lesssim6$\% if $\rbent=1000r_s$ and the disk is warped. 

The power-law indices for the 0.2 keV - 0.5 keV range converges towards $4/3$ with increasing $\rbent$ for broken disks likewise. The broken disk with $\thmax=0.3\pi$ had $\gamma$ varying between $\approx1.15$ ($\approx1.19$ when ignoring $\mu=\pi/2$) and $\approx2.66$ (blue dashed line in the right panel of Fig. \ref{fig:expboth}). Increasing $\rbent$ reduces the peak in $\gamma$ at $\mu=\pi/2+\thmax$, because the energy range $kT_{out}\ll E\lesssim0.3k T_{in}$ is increasingly satisfied by the flat disk region. In addition, $\gamma$ decreases at $\mu=\pi/2$, because an increasingly smaller area is visible. Increasing $\rbent$ to $50 r_s$ and ignoring $\mu=\pi/2$ changes the $\gamma$ range to $\approx0.62-2.18$. Raising $\rbent$ further to $500 r_s$ and $1000 r_s$ makes $\gamma$ increasingly indistinguishable from the multicolor blackbody solution; ignoring $\mu=\pi/2$, $\gamma$ varies between $\approx1.12-1.37$ and $\approx1.34-1.36$ respectively.

Discerning flat and inclined disks through observations will depend on the disk inclination and the observed energy band. A warped disk with $\thmax=0.236\pi$ and $\rbent=500 r_s$ has power-law indices deviating by $\lesssim15$\% from $\gamma=4/3$ for $\mu$ excluding $\pi/2-0.65\pi$, when the innermost region is not visible, and is thus more similar than a warped disk with $\thmax=0.38$ and $\rbent=500 r_s$. The differences between flat and inclined disks are also reduced when contentrating only on the high-energy emission (see Fig. \ref{fig:spectraWarped}, \ref{fig:spectraBroken}). For example, while $\gamma$ of the broken disk with $\thmax=0.3 \pi$ and $\rbent=10r_s$ deviates by a factor of $\approx2$ from $\gamma=4/3$ if $\mu=0.8\pi$, its emission at 8 keV deviates from the flat disk emission only by a factor of $\approx1.16$ for that viewing angle. However, being able to recoup the characteristic power-law index  $\gamma=4/3$ of a multicolor blackbody emission necessitates $\rbent\geq50r_s$.

\section{Discussion}\label{sec:discussion}
The calculations of the spectra of inclined disks 
presented in section \ref{sec:results} are based on simplified accretion disk models (section \ref{sec:methods}). In this section, we discuss the impact of complexities present in physical inclined disks.

We assumed the inclined disks to be infinitely thin, but physical accretion disks extend vertically due to pressure support \citep[e.g.,][]{Blaes2006ApJ...645.1402B,Gu2012ApJ...753..118G}. Self-irradiation will heat the disk and inflate it further, which increases the projected area of the outer disk regions for observers viewing the disk at close to edge-on, and for inner disk regions irradiating them. Accounting for the disk height will therefore enhance the low-energy emission observed at viewing angles close to $\pi/2$, and move the outer disk spectral contribution to higher energies due to increased heating. 


A physical accretion disk may also not radiate everywhere as a blackbody. Blackbody emission requires an optically thick accretion disk, but the density of simulated warped and broken disks varies \citep[e.g.,][]{Nealon2015MNRAS.448.1526N,Liska2019MNRAS.487..550L,Liska2021MNRAS.507..983L}, leading to optical depth variations \citep[e.g.,][]{Liska2023ApJ...944L..48L}. Optically thin regions would emit nonthermal radiation, increasing the deviations from a multicolor blackbody further. 
The thinning of the inclined disk regions could thus not compensate for their geometrical impact. 
Density variations are therefore not expected to eradiacate the differences between warped or broken and flat disk spectra.

Accretion disk dynamics present in inclined disks might also alter their spectra. In broken disks, the interaction between gas in the inclined and the flat plane could cause shocks \citep{Kaaz2023ApJ...955...72K}. The shocked gas will radiate at higher energies and with a spectral distribution following a powerlaw \citep[e.g.,][]{Blandford1978ApJ...221L..29B}. Besides the spectral impact of the powerlaw shape, the shocked gas in the inner disk region also increases the irradiation the outer disk regions. Additional shock heating is thus expected to affect the shape of the inclined disk spectra at high energies, due to the modified distribution of the inner disk emission, and enhance 
the spectral impact of the outer region, due to their stronger irradiation. 

The energy band of the disk spectrum, and thus the band where the power-law index agrees with the multicolor blackbody solution, will furthermore depend on the mass of the black hole and its accretion rate. Decreasing $\dot{M}$ decreases $Q_v$ and $Q_{irr}$, moving the disk spectrum to lower energies. Note that because both heating rates depend on $\dot{M}$, self-irradiation will remain important for large radii of warped disks. The spectrum will also shift to lower energies with increasing black hole mass. However, the spectral shift due to varying $\dot{M}$ and $M$ is minor; the disk temperature due to viscous heating (eq. \ref{eq:qv}) decreases as $\dot{M}^{1/4}$ and $M^{-1/4}$. Therefore, reducing $\dot{M}$ to $0.02\dot{M}_{Edd}$ would still lead to $\gamma\approx4/3$ in the $0.2-0.5$ keV range for a flat disk but to $\gamma$ varying between $0.2-0.96$ for $\mu$ between $0$ and $0.5\pi$ for a warped disk with $\thmax=0.3\pi$. Moving the low-energy disk emission with $\gamma=4/3$ out of the X-ray band would thus require a significant reduction in $\dot{M}$, disagreeing with the Eddington ratios inferred for the soft state of black holes \citep[e.g.,][]{Maccarone2003A&A...409..697M}. Likewise, a significant increase in black hole mass would be necessary to remove the emission less energetic than at the spectral peak from the X-ray band. However, this black hole mass increase would disqualify the black hole from the stellar-mass category \citep[e.g.,][]{Corral2016A&A...587A..61C}, which has been the focus of this study. Hence, for stellar-mass black holes accreting through a misaligned accretion disk with $\rbent=10r_s$ and $\dot{M}\geq0.02\dot{M}_{Edd}$, the discrepancy from a multicolor blackbody should be noticeable.



\section{Conclusion}\label{sec:conclusion}

Misalignments between the black hole spin axis and the angular momentum vector of the accretion flow have been suggested observationally by precessing jets and gravitational wave measurements \citep[e.g,][]{Hjellming1995Natur.375..464H,OShaughnessy2017PhRvL.119a1101O,Cui2023Natur.621..711C}. In simulations, the induced Lense-Thirring precession
warps or breaks the disk depending on the inclination angle. Recent GRMHD simulations of tilted disks showed that disk warping or breaking occurs at $r\lesssim10r_s$ 
\citep[e.g.,][]{Liska2019MNRAS.487..550L,Liska2021MNRAS.507..983L,Liska2022ApJS..263...26L}, while smoothed particle hydrodynamic simulations and analytical calculations place the radius of disk warping or breaking at larger radii \citep[e.g.,][]{Kumar1985MNRAS.213..435K,Nixon2012ApJ...757L..24N}. In this paper, we analytically calculated the thermal emission of warped and broken disks for a variety of disk inclinations and viewing angles (section \ref{sec:methods}), using a similar warp or break radius $\rbent$ as in GRMHD simulations. We find significant differences between inclined and flat disk spectra due to self-irradiation and the projected area of the inclined disk regions (section \ref{sec:results}). The differences are expected to increase when considering a more physical accretion disk setup with e.g., density variations, and should be observable in the X-ray band (section \ref{sec:discussion}). The inclined disk spectra converge towards the multicolor blackbody solution with increasing $\rbent$. While the inferred differences between flat and inclined disk spectra will depend on the disk inclination and the energy band observed, our results suggest that $\rbent$ needs to be $\geq50r_s$ to agree with observations of stellar-mass black holes in the soft state, which can generally be well described with multicolor blackbody spectra. The alternative interpretation is that disk warping or breaking is not common for black hole systems. However, that argument would not be able to reconcile the jet precession observed for the stellar-mass black hole system GRO J1655-40 \citep[][]{Hjellming1995Natur.375..464H}, whose disk emission can be well described as originating from a flat disk \citep[e.g.,][]{Zhang1997ApJ...479..381Z,Sobczak1999ApJ...520..776S,Shafee2006ApJ...636L.113S}.


The comparison between inclined and flat disk spectra presented in this paper might hold implications beyond black holes in the soft state. Low-frequency quasi-periodic oscillations (QPOs) are common in the hard and intermediate state \citep[for a review, see e.g.,][]{Motta2016,Ingram2019}. The preferred model for Type C QPOs is Lense-Thirring precession of the inner hot accretion flow \citep{Ingram2009MNRAS.397L.101I}. However, if the inner hot accretion flow is misaligned enough to cause Lense-Thirring precession, the misalignment should be able to persist in the accretion disk in the soft state. For Type C QPOs to occur in the hard and intermediate state and the misaligned soft state accretion disk to radiate as a multicolor blackbody, $\rbent$ has to vary significantly between spectral states. More work is needed to determine the properties setting the location of the disk warp or break, and how the location changes with the spectral state of the black hole. 



\begin{acknowledgments}
This research was supported in part by the Heising-Simons Foundation, the Simons Foundation, and grants no. NSF PHY-2309135 to the Kavli Institute for Theoretical Physics (KITP).
\end{acknowledgments}

\bibliography{references}{}

\begin{thebibliography}{}
\expandafter\ifx\csname natexlab\endcsname\relax\def\natexlab#1{#1}\fi
\providecommand{\url}[1]{\href{#1}{#1}}
\providecommand{\dodoi}[1]{doi:~\href{http://doi.org/#1}{\nolinkurl{#1}}}
\providecommand{\doeprint}[1]{\href{http://ascl.net/#1}{\nolinkurl{http://ascl.net/#1}}}
\providecommand{\doarXiv}[1]{\href{https://arxiv.org/abs/#1}{\nolinkurl{https://arxiv.org/abs/#1}}}

\bibitem[{{ALMA Partnership} {et~al.}(2015){ALMA Partnership}, {Brogan}, {P{\'e}rez}, {Hunter}, {Dent}, {Hales}, {Hills}, {Corder}, {Fomalont}, {Vlahakis}, {Asaki}, {Barkats}, {Hirota}, {Hodge}, {Impellizzeri}, {Kneissl}, {Liuzzo}, {Lucas}, {Marcelino}, {Matsushita}, {Nakanishi}, {Phillips}, {Richards}, {Toledo}, {Aladro}, {Broguiere}, {Cortes}, {Cortes}, {Espada}, {Galarza}, {Garcia-Appadoo}, {Guzman-Ramirez}, {Humphreys}, {Jung}, {Kameno}, {Laing}, {Leon}, {Marconi}, {Mignano}, {Nikolic}, {Nyman}, {Radiszcz}, {Remijan}, {Rod{\'o}n}, {Sawada}, {Takahashi}, {Tilanus}, {Vila Vilaro}, {Watson}, {Wiklind}, {Akiyama}, {Chapillon}, {de Gregorio-Monsalvo}, {Di Francesco}, {Gueth}, {Kawamura}, {Lee}, {Nguyen Luong}, {Mangum}, {Pietu}, {Sanhueza}, {Saigo}, {Takakuwa}, {Ubach}, {van Kempen}, {Wootten}, {Castro-Carrizo}, {Francke}, {Gallardo}, {Garcia}, {Gonzalez}, {Hill}, {Kaminski}, {Kurono}, {Liu}, {Lopez}, {Morales}, {Plarre}, {Schieven}, {Testi}, {Videla}, {Villard}, {Andreani}, {Hibbard}, \&
  {Tatematsu}}]{Alma2015ApJ...808L...3A}
{ALMA Partnership}, {Brogan}, C.~L., {P{\'e}rez}, L.~M., {et~al.} 2015, \apjl, 808, L3, \dodoi{10.1088/2041-8205/808/1/L3}

\bibitem[{{Balbus} \& {Hawley}(1998)}]{Balbus1998RvMP...70....1B}
{Balbus}, S.~A., \& {Hawley}, J.~F. 1998, Reviews of Modern Physics, 70, 1, \dodoi{10.1103/RevModPhys.70.1}

\bibitem[{{Bardeen} \& {Petterson}(1975)}]{Bardeen1975ApJ...195L..65B}
{Bardeen}, J.~M., \& {Petterson}, J.~A. 1975, \apjl, 195, L65, \dodoi{10.1086/181711}

\bibitem[{{Blaes} {et~al.}(2006){Blaes}, {Davis}, {Hirose}, {Krolik}, \& {Stone}}]{Blaes2006ApJ...645.1402B}
{Blaes}, O.~M., {Davis}, S.~W., {Hirose}, S., {Krolik}, J.~H., \& {Stone}, J.~M. 2006, \apj, 645, 1402, \dodoi{10.1086/503741}

\bibitem[{{Blandford} \& {Ostriker}(1978)}]{Blandford1978ApJ...221L..29B}
{Blandford}, R.~D., \& {Ostriker}, J.~P. 1978, \apjl, 221, L29, \dodoi{10.1086/182658}

\bibitem[{{Blandford} \& {Payne}(1982)}]{Blandford1982MNRAS.199..883B}
{Blandford}, R.~D., \& {Payne}, D.~G. 1982, \mnras, 199, 883, \dodoi{10.1093/mnras/199.4.883}

\bibitem[{{Corral-Santana} {et~al.}(2016){Corral-Santana}, {Casares}, {Mu{\~n}oz-Darias}, {Bauer}, {Mart{\'\i}nez-Pais}, \& {Russell}}]{Corral2016A&A...587A..61C}
{Corral-Santana}, J.~M., {Casares}, J., {Mu{\~n}oz-Darias}, T., {et~al.} 2016, \aap, 587, A61, \dodoi{10.1051/0004-6361/201527130}

\bibitem[{{Cui} {et~al.}(2023){Cui}, {Hada}, {Kawashima}, {Kino}, {Lin}, {Mizuno}, {Ro}, {Honma}, {Yi}, {Yu}, {Park}, {Jiang}, {Shen}, {Kravchenko}, {Algaba}, {Cheng}, {Cho}, {Giovannini}, {Giroletti}, {Jung}, {Lu}, {Niinuma}, {Oh}, {Ohsuga}, {Sawada-Satoh}, {Sohn}, {Takahashi}, {Takamura}, {Tazaki}, {Trippe}, {Wajima}, {Akiyama}, {An}, {Asada}, {Buttaccio}, {Byun}, {Cui}, {Hagiwara}, {Hirota}, {Hodgson}, {Kawaguchi}, {Kim}, {Lee}, {Lee}, {Lee}, {Maccaferri}, {Melis}, {Melnikov}, {Migoni}, {Oh}, {Sugiyama}, {Wang}, {Zhang}, {Chen}, {Hwang}, {Jung}, {Kim}, {Kim}, {Kobayashi}, {Li}, {Li}, {Li}, {Liu}, {Liu}, {Liu}, {Oh}, {Oyama}, {Roh}, {Wang}, {Wang}, {Wang}, {Xia}, {Yan}, {Yeom}, {Yonekura}, {Yuan}, {Zhang}, {Zhao}, \& {Zhong}}]{Cui2023Natur.621..711C}
{Cui}, Y., {Hada}, K., {Kawashima}, T., {et~al.} 2023, \nat, 621, 711, \dodoi{10.1038/s41586-023-06479-6}

\bibitem[{{Done} {et~al.}(2012){Done}, {Davis}, {Jin}, {Blaes}, \& {Ward}}]{Done2012MNRAS.420.1848D}
{Done}, C., {Davis}, S.~W., {Jin}, C., {Blaes}, O., \& {Ward}, M. 2012, \mnras, 420, 1848, \dodoi{10.1111/j.1365-2966.2011.19779.x}

\bibitem[{{Dotani} {et~al.}(1997){Dotani}, {Inoue}, {Mitsuda}, {Nagase}, {Negoro}, {Ueda}, {Makishima}, {Kubota}, {Ebisawa}, \& {Tanaka}}]{Dotani1997ApJ...485L..87D}
{Dotani}, T., {Inoue}, H., {Mitsuda}, K., {et~al.} 1997, \apjl, 485, L87, \dodoi{10.1086/310816}

\bibitem[{{Event Horizon Telescope Collaboration} {et~al.}(2019){Event Horizon Telescope Collaboration}, {Akiyama}, {Alberdi}, {Alef}, {Asada}, {Azulay}, {Baczko}, {Ball}, {Balokovi{\'c}}, {Barrett}, {Bintley}, {Blackburn}, {Boland}, {Bouman}, {Bower}, {Bremer}, {Brinkerink}, {Brissenden}, {Britzen}, {Broderick}, {Broguiere}, {Bronzwaer}, {Byun}, {Carlstrom}, {Chael}, {Chan}, {Chatterjee}, {Chatterjee}, {Chen}, {Chen}, {Cho}, {Christian}, {Conway}, {Cordes}, {Crew}, {Cui}, {Davelaar}, {De Laurentis}, {Deane}, {Dempsey}, {Desvignes}, {Dexter}, {Doeleman}, {Eatough}, {Falcke}, {Fish}, {Fomalont}, {Fraga-Encinas}, {Freeman}, {Friberg}, {Fromm}, {G{\'o}mez}, {Galison}, {Gammie}, {Garc{\'\i}a}, {Gentaz}, {Georgiev}, {Goddi}, {Gold}, {Gu}, {Gurwell}, {Hada}, {Hecht}, {Hesper}, {Ho}, {Ho}, {Honma}, {Huang}, {Huang}, {Hughes}, {Ikeda}, {Inoue}, {Issaoun}, {James}, {Jannuzi}, {Janssen}, {Jeter}, {Jiang}, {Johnson}, {Jorstad}, {Jung}, {Karami}, {Karuppusamy}, {Kawashima}, {Keating}, {Kettenis}, {Kim}, {Kim}, {Kim},
  {Kino}, {Koay}, {Koch}, {Koyama}, {Kramer}, {Kramer}, {Krichbaum}, {Kuo}, {Lauer}, {Lee}, {Li}, {Li}, {Lindqvist}, {Liu}, {Liuzzo}, {Lo}, {Lobanov}, {Loinard}, {Lonsdale}, {Lu}, {MacDonald}, {Mao}, {Markoff}, {Marrone}, {Marscher}, {Mart{\'\i}-Vidal}, {Matsushita}, {Matthews}, {Medeiros}, {Menten}, {Mizuno}, {Mizuno}, {Moran}, {Moriyama}, {Moscibrodzka}, {M{\"u}ller}, {Nagai}, {Nagar}, {Nakamura}, {Narayan}, {Narayanan}, {Natarajan}, {Neri}, {Ni}, {Noutsos}, {Okino}, {Olivares}, {Ortiz-Le{\'o}n}, {Oyama}, {{\"O}zel}, {Palumbo}, {Patel}, {Pen}, {Pesce}, {Pi{\'e}tu}, {Plambeck}, {PopStefanija}, {Porth}, {Prather}, {Preciado-L{\'o}pez}, {Psaltis}, {Pu}, {Ramakrishnan}, {Rao}, {Rawlings}, {Raymond}, {Rezzolla}, {Ripperda}, {Roelofs}, {Rogers}, {Ros}, {Rose}, {Roshanineshat}, {Rottmann}, {Roy}, {Ruszczyk}, {Ryan}, {Rygl}, {S{\'a}nchez}, {S{\'a}nchez-Arguelles}, {Sasada}, {Savolainen}, {Schloerb}, {Schuster}, {Shao}, {Shen}, {Small}, {Sohn}, {SooHoo}, {Tazaki}, {Tiede}, {Tilanus}, {Titus}, {Toma}, {Torne},
  {Trent}, {Trippe}, {Tsuda}, {van Bemmel}, {van Langevelde}, {van Rossum}, {Wagner}, {Wardle}, {Weintroub}, {Wex}, {Wharton}, {Wielgus}, {Wong}, {Wu}, {Young}, {Young}, {Younsi}, {Yuan}, {Yuan}, {Zensus}, {Zhao}, {Zhao}, {Zhu}, {Algaba}, {Allardi}, {Amestica}, {Anczarski}, {Bach}, {Baganoff}, {Beaudoin}, {Benson}, {Berthold}, {Blanchard}, {Blundell}, {Bustamente}, {Cappallo}, {Castillo-Dom{\'\i}nguez}, {Chang}, {Chang}, {Chang}, {Chen}, {Chilson}, {Chuter}, {C{\'o}rdova Rosado}, {Coulson}, {Crawford}, {Crowley}, {David}, {Derome}, {Dexter}, {Dornbusch}, {Dudevoir}, {Dzib}, {Eckart}, {Eckert}, {Erickson}, {Everett}, {Faber}, {Farah}, {Fath}, {Folkers}, {Forbes}, {Freund}, {G{\'o}mez-Ruiz}, {Gale}, {Gao}, {Geertsema}, {Graham}, {Greer}, {Grosslein}, {Gueth}, {Haggard}, {Halverson}, {Han}, {Han}, {Hao}, {Hasegawa}, {Henning}, {Hern{\'a}ndez-G{\'o}mez}, {Herrero-Illana}, {Heyminck}, {Hirota}, {Hoge}, {Huang}, {Impellizzeri}, {Jiang}, {Kamble}, {Keisler}, {Kimura}, {Kono}, {Kubo}, {Kuroda}, {Lacasse}, {Laing},
  {Leitch}, {Li}, {Lin}, {Liu}, {Liu}, {Lu}, {Marson}, {Martin-Cocher}, {Massingill}, {Matulonis}, {McColl}, {McWhirter}, {Messias}, {Meyer-Zhao}, {Michalik}, {Monta{\~n}a}, {Montgomerie}, {Mora-Klein}, {Muders}, {Nadolski}, {Navarro}, {Neilsen}, {Nguyen}, {Nishioka}, {Norton}, {Nowak}, {Nystrom}, {Ogawa}, {Oshiro}, {Oyama}, {Parsons}, {Paine}, {Pe{\~n}alver}, {Phillips}, {Poirier}, {Pradel}, {Primiani}, {Raffin}, {Rahlin}, {Reiland}, {Risacher}, {Ruiz}, {S{\'a}ez-Mada{\'\i}n}, {Sassella}, {Schellart}, {Shaw}, {Silva}, {Shiokawa}, {Smith}, {Snow}, {Souccar}, {Sousa}, {Sridharan}, {Srinivasan}, {Stahm}, {Stark}, {Story}, {Timmer}, {Vertatschitsch}, {Walther}, {Wei}, {Whitehorn}, {Whitney}, {Woody}, {Wouterloot}, {Wright}, {Yamaguchi}, {Yu}, {Zeballos}, {Zhang}, \& {Ziurys}}]{EHT2019ApJ...875L...1E}
{Event Horizon Telescope Collaboration}, {Akiyama}, K., {Alberdi}, A., {et~al.} 2019, \apjl, 875, L1, \dodoi{10.3847/2041-8213/ab0ec7}

\bibitem[{{Fragile} \& {Liska}(2024)}]{Fragile2024arXiv240410052F}
{Fragile}, P.~C., \& {Liska}, M. 2024, arXiv e-prints, arXiv:2404.10052, \dodoi{10.48550/arXiv.2404.10052}

\bibitem[{{Frontera} {et~al.}(2001){Frontera}, {Palazzi}, {Zdziarski}, {Haardt}, {Perola}, {Chiappetti}, {Cusumano}, {Dal Fiume}, {Del Sordo}, {Orlandini}, {Parmar}, {Piro}, {Santangelo}, {Segreto}, {Treves}, \& {Trifoglio}}]{Frontera2001ApJ...546.1027F}
{Frontera}, F., {Palazzi}, E., {Zdziarski}, A.~A., {et~al.} 2001, \apj, 546, 1027, \dodoi{10.1086/318304}

\bibitem[{{Fukue}(1992)}]{Fukue1992PASJ...44..663F}
{Fukue}, J. 1992, PASJ, 44, 663

\bibitem[{{Gerosa} {et~al.}(2013){Gerosa}, {Kesden}, {Berti}, {O'Shaughnessy}, \& {Sperhake}}]{Gerosa2013PhRvD..87j4028G}
{Gerosa}, D., {Kesden}, M., {Berti}, E., {O'Shaughnessy}, R., \& {Sperhake}, U. 2013, \prd, 87, 104028, \dodoi{10.1103/PhysRevD.87.104028}

\bibitem[{{Gierli{\'n}ski} \& {Done}(2004)}]{Gierlinski2004MNRAS.347..885G}
{Gierli{\'n}ski}, M., \& {Done}, C. 2004, \mnras, 347, 885, \dodoi{10.1111/j.1365-2966.2004.07266.x}

\bibitem[{{Gu}(2012)}]{Gu2012ApJ...753..118G}
{Gu}, W.-M. 2012, \apj, 753, 118, \dodoi{10.1088/0004-637X/753/2/118}

\bibitem[{{Hjellming} \& {Rupen}(1995)}]{Hjellming1995Natur.375..464H}
{Hjellming}, R.~M., \& {Rupen}, M.~P. 1995, \nat, 375, 464, \dodoi{10.1038/375464a0}

\bibitem[{{Ingram} {et~al.}(2009){Ingram}, {Done}, \& {Fragile}}]{Ingram2009MNRAS.397L.101I}
{Ingram}, A., {Done}, C., \& {Fragile}, P.~C. 2009, \mnras, 397, L101, \dodoi{10.1111/j.1745-3933.2009.00693.x}

\bibitem[{{Ingram} \& {Motta}(2019)}]{Ingram2019}
{Ingram}, A.~R., \& {Motta}, S.~E. 2019, \nar, 85, 101524, \dodoi{10.1016/j.newar.2020.101524}

\bibitem[{{Kaaz} {et~al.}(2023){Kaaz}, {Liska}, {Jacquemin-Ide}, {Andalman}, {Musoke}, {Tchekhovskoy}, \& {Porth}}]{Kaaz2023ApJ...955...72K}
{Kaaz}, N., {Liska}, M. T.~P., {Jacquemin-Ide}, J., {et~al.} 2023, \apj, 955, 72, \dodoi{10.3847/1538-4357/ace051}

\bibitem[{{Kalogera}(2000)}]{Kalogera2000ApJ...541..319K}
{Kalogera}, V. 2000, \apj, 541, 319, \dodoi{10.1086/309400}

\bibitem[{{King} \& {Pringle}(2006)}]{King2006MNRAS.373L..90K}
{King}, A.~R., \& {Pringle}, J.~E. 2006, \mnras, 373, L90, \dodoi{10.1111/j.1745-3933.2006.00249.x}

\bibitem[{{Kumar} \& {Pringle}(1985)}]{Kumar1985MNRAS.213..435K}
{Kumar}, S., \& {Pringle}, J.~E. 1985, \mnras, 213, 435, \dodoi{10.1093/mnras/213.3.435}

\bibitem[{{Liska} {et~al.}(2021){Liska}, {Hesp}, {Tchekhovskoy}, {Ingram}, {van der Klis}, {Markoff}, \& {Van Moer}}]{Liska2021MNRAS.507..983L}
{Liska}, M., {Hesp}, C., {Tchekhovskoy}, A., {et~al.} 2021, \mnras, 507, 983, \dodoi{10.1093/mnras/staa099}

\bibitem[{{Liska} {et~al.}(2019){Liska}, {Tchekhovskoy}, {Ingram}, \& {van der Klis}}]{Liska2019MNRAS.487..550L}
{Liska}, M., {Tchekhovskoy}, A., {Ingram}, A., \& {van der Klis}, M. 2019, \mnras, 487, 550, \dodoi{10.1093/mnras/stz834}

\bibitem[{{Liska} {et~al.}(2023){Liska}, {Kaaz}, {Musoke}, {Tchekhovskoy}, \& {Porth}}]{Liska2023ApJ...944L..48L}
{Liska}, M.~T.~P., {Kaaz}, N., {Musoke}, G., {Tchekhovskoy}, A., \& {Porth}, O. 2023, \apjl, 944, L48, \dodoi{10.3847/2041-8213/acb6f4}

\bibitem[{{Liska} {et~al.}(2022){Liska}, {Chatterjee}, {Issa}, {Yoon}, {Kaaz}, {Tchekhovskoy}, {van Eijnatten}, {Musoke}, {Hesp}, {Rohoza}, {Markoff}, {Ingram}, \& {van der Klis}}]{Liska2022ApJS..263...26L}
{Liska}, M.~T.~P., {Chatterjee}, K., {Issa}, D., {et~al.} 2022, \apjs, 263, 26, \dodoi{10.3847/1538-4365/ac9966}

\bibitem[{{Maccarone}(2003)}]{Maccarone2003A&A...409..697M}
{Maccarone}, T.~J. 2003, \aap, 409, 697, \dodoi{10.1051/0004-6361:20031146}

\bibitem[{{Makishima} {et~al.}(1986){Makishima}, {Maejima}, {Mitsuda}, {Bradt}, {Remillard}, {Tuohy}, {Hoshi}, \& {Nakagawa}}]{Makishima1986ApJ...308..635M}
{Makishima}, K., {Maejima}, Y., {Mitsuda}, K., {et~al.} 1986, \apj, 308, 635, \dodoi{10.1086/164534}

\bibitem[{{Mall} {et~al.}(2024){Mall}, {Liu}, {Bambi}, {Steiner}, \& {Garc{\'\i}a}}]{Mall2024MNRAS.52712053M}
{Mall}, G., {Liu}, H., {Bambi}, C., {Steiner}, J.~F., \& {Garc{\'\i}a}, J.~A. 2024, \mnras, 527, 12053, \dodoi{10.1093/mnras/stad3933}

\bibitem[{{Motta}(2016)}]{Motta2016}
{Motta}, S.~E. 2016, Astronomische Nachrichten, 337, 398, \dodoi{10.1002/asna.201612320}

\bibitem[{{Mu{\~n}oz-Darias} {et~al.}(2019){Mu{\~n}oz-Darias}, {Jim{\'e}nez-Ibarra}, {Panizo-Espinar}, {Casares}, {Mata S{\'a}nchez}, {Ponti}, {Fender}, {Buckley}, {Garnavich}, {Torres}, {Armas Padilla}, {Charles}, {Corral-Santana}, {Kajava}, {Kotze}, {Littlefield}, {S{\'a}nchez-Sierras}, {Steeghs}, \& {Thomas}}]{MunozDarias2019ApJ...879L...4M}
{Mu{\~n}oz-Darias}, T., {Jim{\'e}nez-Ibarra}, F., {Panizo-Espinar}, G., {et~al.} 2019, \apjl, 879, L4, \dodoi{10.3847/2041-8213/ab2768}

\bibitem[{{Nealon} {et~al.}(2015){Nealon}, {Price}, \& {Nixon}}]{Nealon2015MNRAS.448.1526N}
{Nealon}, R., {Price}, D.~J., \& {Nixon}, C.~J. 2015, \mnras, 448, 1526, \dodoi{10.1093/mnras/stv014}

\bibitem[{{Nelson} \& {Papaloizou}(2000)}]{Nelson2000MNRAS.315..570N}
{Nelson}, R.~P., \& {Papaloizou}, J. C.~B. 2000, \mnras, 315, 570, \dodoi{10.1046/j.1365-8711.2000.03478.x}

\bibitem[{{Nixon} {et~al.}(2012){Nixon}, {King}, {Price}, \& {Frank}}]{Nixon2012ApJ...757L..24N}
{Nixon}, C., {King}, A., {Price}, D., \& {Frank}, J. 2012, \apjl, 757, L24, \dodoi{10.1088/2041-8205/757/2/L24}

\bibitem[{{Orosz} \& {Bailyn}(1997)}]{Orosz1997ApJ...477..876O}
{Orosz}, J.~A., \& {Bailyn}, C.~D. 1997, \apj, 477, 876, \dodoi{10.1086/303741}

\bibitem[{{O'Shaughnessy} {et~al.}(2017){O'Shaughnessy}, {Gerosa}, \& {Wysocki}}]{OShaughnessy2017PhRvL.119a1101O}
{O'Shaughnessy}, R., {Gerosa}, D., \& {Wysocki}, D. 2017, \prl, 119, 011101, \dodoi{10.1103/PhysRevLett.119.011101}

\bibitem[{{Scheuer} \& {Feiler}(1996)}]{Scheuer1996MNRAS.282..291S}
{Scheuer}, P.~A.~G., \& {Feiler}, R. 1996, \mnras, 282, 291, \dodoi{10.1093/mnras/282.1.291}

\bibitem[{{Shafee} {et~al.}(2006){Shafee}, {McClintock}, {Narayan}, {Davis}, {Li}, \& {Remillard}}]{Shafee2006ApJ...636L.113S}
{Shafee}, R., {McClintock}, J.~E., {Narayan}, R., {et~al.} 2006, \apjl, 636, L113, \dodoi{10.1086/498938}

\bibitem[{{Shakura} \& {Sunyaev}(1973)}]{Shakura1973A&A....24..337S}
{Shakura}, N.~I., \& {Sunyaev}, R.~A. 1973, AAP, 24, 337

\bibitem[{{Sobczak} {et~al.}(1999){Sobczak}, {McClintock}, {Remillard}, {Bailyn}, \& {Orosz}}]{Sobczak1999ApJ...520..776S}
{Sobczak}, G.~J., {McClintock}, J.~E., {Remillard}, R.~A., {Bailyn}, C.~D., \& {Orosz}, J.~A. 1999, \apj, 520, 776, \dodoi{10.1086/307474}

\bibitem[{{Spruit}(1989)}]{Spruit1989ASIC..290..325S}
{Spruit}, H.~C. 1989, in NATO Advanced Study Institute (ASI) Series C, Vol. 290, Theory of Accretion Disks, ed. F.~{Meyer}, 325--340

\bibitem[{{Taverna} {et~al.}(2020){Taverna}, {Zhang}, {Dov{\v{c}}iak}, {Bianchi}, {Bursa}, {Karas}, \& {Matt}}]{Taverna2020MNRAS.493.4960T}
{Taverna}, R., {Zhang}, W., {Dov{\v{c}}iak}, M., {et~al.} 2020, \mnras, 493, 4960, \dodoi{10.1093/mnras/staa598}

\bibitem[{{van den Eijnden} {et~al.}(2020){van den Eijnden}, {Degenaar}, {Ludlam}, {Parikh}, {Miller}, {Wijnands}, {Gendreau}, {Arzoumanian}, {Chakrabarty}, \& {Bult}}]{Eijnden2020MNRAS.493.1318V}
{van den Eijnden}, J., {Degenaar}, N., {Ludlam}, R.~M., {et~al.} 2020, \mnras, 493, 1318, \dodoi{10.1093/mnras/staa423}

\bibitem[{{Williams} \& {Cieza}(2011)}]{Williams2011ARA&A..49...67W}
{Williams}, J.~P., \& {Cieza}, L.~A. 2011, \araa, 49, 67, \dodoi{10.1146/annurev-astro-081710-102548}

\bibitem[{{Zhang} {et~al.}(1997){Zhang}, {Ebisawa}, {Sunyaev}, {Ueda}, {Harmon}, {Sazonov}, {Fishman}, {Inoue}, {Paciesas}, \& {Takahash}}]{Zhang1997ApJ...479..381Z}
{Zhang}, S.~N., {Ebisawa}, K., {Sunyaev}, R., {et~al.} 1997, \apj, 479, 381, \dodoi{10.1086/303870}

\end{thebibliography}
\bibliographystyle{aasjournal}
\end{document}